\documentstyle[12pt,epsfig]{article}
                                                                                
\title{                                                                         
{\vspace{-3cm} \normalsize                                                      
\hfill \parbox{30mm}{DESY 95-192}}\\[25mm]                                      
An algorithm for gluinos on the lattice          \\[8mm]}
\author{I. Montvay                               \\
Deutsches Elektronen-Synchrotron DESY,           \\
Notkestr.\,85, D-22603 Hamburg, Germany}                                        
                                                                                
\date{October, 1995}
                                                                                
\newcommand{\be}{\begin{equation}}                                              
\newcommand{\ee}{\end{equation}}                                                
\newcommand{\half}{\frac{1}{2}}                                                 
                                                  
\newcommand{\lar}{\leftarrow}
\newcommand{\LCB}{\raisebox{-0.3ex}{\mbox{\LARGE$\left\{\right.$}}}
\newcommand{\RCB}{\raisebox{-0.3ex}{\mbox{\LARGE$\left.\right\}$}}}
                                                                                
\begin{document}                                                                
\maketitle                                                                      
                                                                                
\begin{abstract} \normalsize                                                    
 L\"uscher's local bosonic algorithm for Monte Carlo simulations of
 quantum field theories with fermions is applied to the simulation
 of a possibly supersymmetric Yang-Mills theory with a Majorana
 fermion in the adjoint representation.
 Combined with a correction step in a two-step polynomial
 approximation scheme, the obtained algorithm seems to be promising
 and could be competitive with more conventional algorithms based on
 discretized classical (``molecular dyna\-mics'') equations of motion.
 The application of the considered polynomial approximation scheme to
 optimized hopping parameter expansions is also discussed.
\end{abstract}                                                                  
\hspace{1cm}                                                                    
                                                                                
\section{Introduction}                                 \label{sec1}
 Among quantum field theories the supersymmetric ones play a very
 distinguished r\^ole: they have much less free parameters than a
 general renormalizable quantum field theory with the same set of
 fields, and show remarkable non-renormalization and finiteness
 properties (for general references see, for instance, \cite{SUSY}).
 Even more special are supersymmetric Yang Mills (SYM) theories.
 The Yang Mills theory with a Majorana fermion in the adjoint
 representation is automatically $N=1$ supersymmetric in the massless
 case, at least according to perturbation theory.
 The supersymmetry constraints in $N=2$ SYM theory can be exploited to
 determine, under some reasonable assumptions, its exact solution in
 the low energy limit \cite{SEIWIT1}.

 The non-perturbative properties of supesymmetric quantum gauge field
 theories enjoyed particular interest already in the 80's (see the
 review \cite{AKMRV} and references therein).
 After the recent beautiful exact results of Seiberg and Witten
 \cite{SEIWIT1,SEIWIT2} there is a revived interest because one
 can hope to understand better on these examples some general
 non-perturbative phenomana as confinement and chiral symmetry
 breaking.

 It is obvious that numerical simulations for studying non-perturbative
 properties of supersymmetric quantum field theories would be both
 interesting and desirable.
 In this way the assumptions beyond the exact solutions could, in
 principle, be checked and the results perhaps extended.
 Nevertheless, up to now it turned out to be impossible to find a 
 lattice formulation with exact supersymmetry.
 The only possibility for reconciling lattice regularization with
 supersymmetry seems to be to formulate a more general
 non-supersymmetric model and recover supersymmetry in the continuum
 limit as a result of fine-tuning the parameters.
 This approach was pioneered by Curci and Veneziano \cite{CURVEN} in
 the simple case of $N=1$ SYM, where there is only one
 parameter to tune (namely, the Majorana fermion mass).
 More general cases can be expected to be dealt with similarly, as
 shown, for instance, in the case of $N=2$ SYM in ref.\ \cite{N=2}.
 This way of ``embedding'' supersymmetric theories in more general
 non-supersymmetric ones is natural since in Nature supersymmetry
 is also broken.
 Besides, as argued recently \cite{ASPY}, many of the ``exotic'' 
 dynamical features of supersymmetric theories survive, if small
 symmetry breaking terms are added.

 Since supersymmetry connects bosons and fermions, a numerical
 simulation also involves the numerical approximation of fermionic
 Grassmann integrals.
 This is a notoriously difficult problem, which makes numerical
 simulations of e.~g.\ QCD with dynamical fermions quite hard.
 Nevertheless, there are several known fermion algorithms which
 do the job for QCD, for instance, the popular Hybrid Monte Carlo
 algorithm \cite{HMC}.
 This is not directly applicable in supersymmetric models where
 Majorana fermions are important.
 Nevertheless, some related algorithms based on discretized classical
 ``molecular dynamics'' equations of motions, as for instance the one
 in ref. \cite{HCL}, can work.
 However, recently L\"uscher proposed a conceptually rather different
 local bosonic algorithm \cite{LUSCHER}, which is an alternative
 for QCD \cite{BJJLSS}, and can also be extended to supersymmetric
 cases.
 As recently pointed out by Bori\c{c}i and de Forcrand \cite{BORFOR},
 this algorithm gains on attractivity, if it is combined with a
 ``noisy correction'' step as proposed in the early days of fermion
 algorithm developments \cite{KENKUT}.
 This local bosonic formulation and its combination with a noisy
 correction step is the basis of the fermion algorithm investigated
 in the present paper.
 The formulation and tests were done in $N=1$ SYM with SU(2) gauge
 group, but the methods used are obviously more general: they can be
 applied in many supersymmetric models with $N=1$ and $N>1$
 supersymmetry and, of course, also in other models containing
 fermions and not related to supersymmetry.

 The plan of this paper is the following: In the next section
 the lattice formulation of models with Majorana fermions is
 discussed on the example of $N=1$ SYM.
 In section~\ref{sec3} the optimized polynomial approximation
 scheme is introduced, which is then used in the local bosonic
 algorithms defined and tested in section~\ref{sec4}.
 Section~\ref{sec5} is devoted to the discussion of the applications
 of the optimized polynomial approximations to hopping parameter
 expansions.
 A possible fermion algorithm based on optimized numerical hopping
 parameter expansion is also briefly discussed there.
 Finally, section \ref{sec6} contains some concluding remarks.
                                                                                
\section{Majorana fermions on the lattice}             \label{sec2}
 In order to define the path integral for a Yang Mills theory with
 Majorana fermions in the adjoint representation, let us first
 consider the same theory with Dirac fermions.
 The lattice action in case of Wilson's lattice fermion formulation
 has been considered in \cite{N=2}.
 If the Grassmanian fermion fields in the adjoint representation are
 denoted by $\psi^r_x$ and $\overline{\psi}^r_x$, with $r$ being the
 adjoint representation index ($r=1,..,N_c^2-1$ for SU($N_c$) ), then
 the fermionic part of the lattice action is: 
\be  \label{eq01}
S_f = \sum_x \LCB \overline{\psi}_x^r\psi_x^r
-K \sum_{\mu=1}^4 \left[
\overline{\psi}_{x+\hat{\mu}}^r V_{rs,x\mu}(1+\gamma_\mu)\psi_x^s
+\overline{\psi}_x^r V_{rs,x\mu}^T (1-\gamma_\mu)
\psi_{x+\hat{\mu}}^s \right] \RCB \ .
\ee
 Here $K$ is the hopping parameter, the irrelevant Wilson parameter
 removing the fermion doublers in the continuum limit is fixed to
 $r=1$, and the matrix for the gauge-field link in the adjoint
 representation is defined as
\be  \label{eq02}
V_{rs,x\mu} \equiv V_{rs,x\mu}[U] \equiv
2 {\rm Tr}(U_{x\mu}^\dagger T_r U_{x\mu} T_s)
= V_{rs,x\mu}^* =V_{rs,x\mu}^{-1T} \ .
\ee
 The generators $T_r \equiv \half \lambda_r$ satisfy the usual
 normalization ${\rm Tr\,}(\lambda_r\lambda_s)=\half$.
 In case of SU(2) ($N_c=2$) we have, of course,
 $T_r \equiv \half \tau_r$ with the isospin Pauli-matrices $\tau_r$.
 The normalization of the fermion fields in (\ref{eq01}) is the
 usual one for numerical simulations.
 The full lattice action is the sum of the pure gauge part and
 fermionic part: 
\be  \label{eq03}
S = S_g + S_f \ .
\ee
 The standard Wilson action for the SU($N_c$) gauge field $S_g$
 is a sum over the plaquettes
\be  \label{eq04}
S_g  =   \beta \sum_{pl}                                                  
\left( 1 - \frac{1}{N_c} {\rm Re\,Tr\,} U_{pl} \right) \ ,   
\ee
 with the bare gauge coupling given by $\beta \equiv 2N_c/g^2$.

 In order to obtain the lattice formulation of a theory with
 Majorana fermions, let us introduce the Majorana field components
\be  \label{eq05}
\Psi^{(1)} \equiv \frac{1}{\sqrt{2}} ( \psi + C\overline{\psi}^T)
\ , \hspace{2em}
\Psi^{(2)} \equiv \frac{i}{\sqrt{2}} (-\psi + C\overline{\psi}^T)
\ee
 with the charge conjugation matrix $C$.
 These satisfy the Majorana condition
\be  \label{eq06}
\overline{\Psi}^{(j)} = \Psi^{(j)T} C
\hspace{3em} (j=1,2)  \ . 
\ee
 The inverse relation of (\ref{eq05}) is
\be  \label{eq07}
\psi = \frac{1}{\sqrt{2}} (\Psi^{(1)} + i\Psi^{(2)})
\ , \hspace{2em}
\psi_c \equiv C\overline{\psi}^T =
\frac{1}{\sqrt{2}} (\Psi^{(1)} - i\Psi^{(2)}) \ .
\ee
 In terms of the two Majorana fields the fermion action $S_f$ in
 eq.\ (\ref{eq01}) can be written as
\be  \label{eq08}
S_f = \half \sum_x \sum_{j=1}^2 \LCB
\overline{\Psi}_x^{(j)r}\Psi_x^{(j)r} 
-K \sum_{\mu=1}^4 \left[
\overline{\Psi}_{x+\hat{\mu}}^{(j)r} V_{rs,x\mu}
(1+\gamma_\mu)\Psi_x^{(j)s}
+\overline{\Psi}_x^{(j)r} V_{rs,x\mu}^T (1-\gamma_\mu)
\Psi_{x+\hat{\mu}}^{(j)s} \right] \RCB \ .
\ee

 For later purposes it is convenient to introduce the {\em fermion
 matrix}
\be  \label{eq09}
Q_{yv,xu} \equiv Q_{yv,xu}[U] \equiv
\delta_{yx}\delta_{vu} - K \sum_{\mu=1}^4 \left[
\delta_{y,x+\hat{\mu}}(1+\gamma_\mu) V_{vu,x\mu} +
\delta_{y+\hat{\mu},x}(1-\gamma_\mu) V^T_{vu,y\mu} \right] \ .
\ee
 Here, as usual, $\hat{\mu}$ denotes the unit vector in direction
 $\mu$.
 In terms of $Q$ we have
\be  \label{eq10}
S_f = \sum_{xu,yv} \overline{\psi}^v_y Q_{yv,xu} \psi^u_x
= \half\sum_{j=1}^2
\sum_{xu,yv} \overline{\Psi}^{(j)v}_y Q_{yv,xu} \Psi^{(j)u}_x \ ,
\ee
 and the fermionic path integral can be written as
\be  \label{eq11}
\int [d\overline{\psi} d\psi] e^{-S_f} = 
\int [d\overline{\psi} d\psi] e^{-\overline{\psi} Q \psi} = \det Q
= \prod_{j=1}^2 \int [d\Psi^{(j)}] 
e^{ -\half\overline{\Psi}^{(j)}Q\Psi^{(j)} } \ .
\ee
 Note that for Majorana fields the path integral involves only
 $[d\Psi^{(j)}]$ because of the Majorana condition in (\ref{eq06}).
 This implies for $\Psi \equiv \Psi^{(1)}$ or $\Psi \equiv \Psi^{(2)}$
\be  \label{eq12}
\int [d\Psi] e^{ -\half\overline{\Psi} Q \Psi }
= \pm \sqrt{\det Q} \ .
\ee
 For a given gauge field the sign can be taken by definition
 to be positive, but for different gauge fields one has to care about
 continuity.
 The sign convention can be fixed, for instance, at the trivial
 gauge field $U_{x\mu} \equiv 1$ to be positive.
 Then the positive sign stays by continuity until one reaches gauge
 fields with $\det Q[U]=0$.
 
 The squre root of the determinant in eq.\ (\ref{eq12}) is a
 {\em Pfaffian}.
 This can be defined for a general complex antisymmetric matrix
 $M_{\alpha\beta}=-M_{\beta\alpha}$ with an even number of dimensions
 ($1 \leq \alpha,\beta \leq 2N$) by a Grassmann integral as
\be  \label{eq13}
{\rm pf}(M) \equiv
\int [d\phi] e^{-\phi_\alpha M_{\alpha\beta} \phi_\beta}
= \frac{1}{N!} \epsilon_{\alpha_1\beta_1 \ldots \alpha_N\beta_N}
M_{\alpha_1\beta_1} \ldots M_{\alpha_N\beta_N} \ .
\ee
 Here, of course, $[d\phi] \equiv d\phi_{2N} \ldots d\phi_1$, and 
 $\epsilon$ is the totally antisymmetric unit tensor.
 Using the above trick with doubling the number of Grassmann
 variables one can easily show that
\be  \label{eq14}
\left[{\rm pf}(\half M)\right]^2 = \det M \ .
\ee
 This establishes the connection between Pfaffians and determinants.
 According to eq.\ (\ref{eq06}), in case of the above fermion action
 the antisymmetric matrix is $\bar{Q} \equiv CQ$.

 It is now clear that the fermion action for a Majorana fermion in the
 adjoint representation $\Psi^r_x$ can be defined by
\be  \label{eq15}
S_f \equiv \half \overline{\Psi} Q \Psi \equiv 
\half \sum_x \LCB \overline{\Psi}_x^r\Psi_x^r
-K \sum_{\mu=1}^4 \left[
\overline{\Psi}_{x+\hat{\mu}}^r V_{rs,x\mu}(1+\gamma_\mu)\Psi_x^s
+\overline{\Psi}_x^r V_{rs,x\mu}^T (1-\gamma_\mu)
\Psi_{x+\hat{\mu}}^s \right] \RCB \ .
\ee
 The path integral over $\Psi$ is defined by the Pfaffian
 ${\rm pf}(\half CQ)=\sqrt{\det(CQ)}=\sqrt{\det(Q)}$.
 In this way the Majorana nature of $\Psi$ implies that one has
 to take the square root of the usual fermion determinant
 \cite{CURVEN}.
 Note that the Pfaffian defined by (\ref{eq13}) assigns a unique sign
 to the path integral for Majorana fermions, therefore there is no
 sign ambiguity.
 For positive determinant the Pfaffian is real, but if the
 determinant is negative then the Pfaffian has to be pure imaginary.

 Expectation values of Majorana fermion fields can be calculated
 as follows.
 For the Dirac fermion fields $\psi,\overline{\psi}$ we have, as is
 well known,
$$
\left\langle \psi_{y_1} \overline{\psi}_{x_1}
             \psi_{y_2} \overline{\psi}_{x_2}
\cdots       \psi_{y_n} \overline{\psi}_{x_n} \right\rangle =
Z^{-1} \int [d U] e^{-S_g[U]} \det Q[U]
$$
\be \label{eq16}
\cdot \sum_{z_1 \cdots z_n}
\epsilon^{z_1 z_2 \cdots z_n}_{y_1 y_2 \cdots y_n}
Q[U]^{-1}_{z_1x_1} Q[U]^{-1}_{z_2x_2} \cdots Q[U]^{-1}_{z_nx_n}
\ ,
\ee
with the antisymmetrizing unit tensor $\epsilon$ and
\be \label{eq17}
Z \equiv \int [d U] e^{-S_g[U]} \det Q[U] \ .
\ee
 In order to express the expectation value of Majorana fermion fields
 by the matrix elements of the propagator $Q[U]^{-1}$, one can use
 again the doubling trick: one can identify, for instance,
 $\Psi \equiv \Psi^{(1)}$ and introduce another Majorana field
 $\Psi^{(2)}$, in order to obtain a Dirac field.
 Then from eqs.\ (\ref{eq05}) and (\ref{eq16}) we obtain
$$
\left\langle \Psi_{y_1} \overline{\Psi}_{x_1}
             \Psi_{y_2} \overline{\Psi}_{x_2}
\cdots       \Psi_{y_n} \overline{\Psi}_{x_n} \right\rangle =
Z_M^{-1} \int [d U] e^{-S_g[U]} \sqrt{\det Q[U]}
$$
\be \label{eq18}
\cdot (\det Q[U])^{-1} \int [d\overline{\psi} d\psi] 
e^{-\overline{\psi} Q \psi} 
(\overline{\psi}_{x_1}+C\psi^T_{x_1})  
(\overline{\psi}_{x_2}+C\psi^T_{x_2}) \cdots 
(\overline{\psi}_{x_n}+C\psi^T_{x_n}) 2^{-n} 
\ ,
\ee
 where now
\be \label{eq19}
Z_M \equiv \int [d U] e^{-S_g[U]} \sqrt{\det Q[U]} \ .
\ee
 In particular, for $n=1$ we have with $Q \equiv Q[U]$
\be \label{eq20}
\left\langle \Psi_y \overline{\Psi}_x \right\rangle =
Z_M^{-1} \int [d U] e^{-S_g[U]} \sqrt{\det Q[U]}\;
\half \left\{ Q^{-1}_{yx} + C^{-1}Q^{-1}_{xy}C \right\} \ .
\ee
 The important case $n=2$ can be expressed by six terms:
$$
\left\langle \Psi_{y_2} \overline{\Psi}_{x_2}
             \Psi_{y_1} \overline{\Psi}_{x_1} \right\rangle 
= Z_M^{-1} \int [d U] e^{-S_g[U]} \sqrt{\det Q[U]}
$$
$$
\cdot \frac{1}{4} \sum_{z_1z_2} \left\{ \epsilon^{z_1z_2}_{y_1y_2}
Q^{-1}_{z_1x_1} Q^{-1}_{z_2x_2} +
C_{x_1}^{-1}\epsilon^{z_1z_2}_{x_1y_2} 
Q^{-1}_{z_1y_1} Q^{-1}_{z_2x_2}C_{y_1} + 
C_{x_2}^{-1}\epsilon^{z_1z_2}_{y_1x_2} 
Q^{-1}_{z_1x_1} Q^{-1}_{z_2y_2}C_{y_2} 
\right. 
$$
\be \label{eq21}
\left. + 
C_{x_1}^{-1}C_{x_2}^{-1} \epsilon^{z_1z_2}_{x_1x_2} 
Q^{-1}_{z_1y_1} Q^{-1}_{z_2y_2} C_{y_1}C_{y_2} -
C_{x_1}^{-1}\epsilon^{z_1z_2}_{y_1x_1} 
Q^{-1}_{z_1x_2} Q^{-1}_{z_2y_2}C_{y_2} - 
C_{x_2}^{-1}\epsilon^{z_1z_2}_{y_2x_2} 
Q^{-1}_{z_1x_1} Q^{-1}_{z_2y_1}C_{y_1} \right\} \ .
\ee
 The indices on the charge conjugation matrix $C$ show how the Dirac
 indices have to be contracted.

 Since the path integral over Majorana fermion fields is defined by
 a Pfaffian, the sign in eq.\ (\ref{eq12}) is uniquely determined.
 It is easy to show that the fermion matrix in (\ref{eq09}) satisfies
\be \label{eq22}
Q^\dagger = \gamma_5 Q \gamma_5 \ , 
\ee
 therefore $\det(Q)$ is always real.
 In principle, in numerical simulations one can take into account the
 phase of the Pfaffian by performing the Monte Carlo integration with
 the positive square root $+\sqrt{|\det Q|}$, and correct for the phase
 by putting it into the expectation values.
 Although this solves the phase problem in principle, in practice there
 is still a problem because the numerical evaluation of the Pfaffian
 from the formula (\ref{eq13}) is rather cumbersome.
 The only practical possibility in a Monte Carlo process is to monitor
 the lowest eigenvalues of $Q$ and/or explicitly calculate $\det Q$,
 in order to make it probable that $\det Q$ does not cross zero,
 where the question of the sign change in eq.~(\ref{eq12}) and the
 problem of the non-trivial phase of the Pfaffian emerge.
 In what follows it will be assumed that one can restrict the
 Monte Carlo simulation to gauge configurations where
 ${\rm pf}(\half CQ)=+\sqrt{|\det(Q)|}$.

\section{Optimized polynomial approximations}          \label{sec3}
 An important ingredient of L\"uscher's fermion algorithm
 \cite{LUSCHER} is the polynomial approximation of the function $1/x$
 in some positive interval $x \in [\epsilon,\lambda]$
 $(0 < \epsilon < \lambda < \infty)$.
 This can be achieved by Chebyshev polynomials, which minimize
 the maximum relative error (``infinity-norm'').
 Here we shall follow another approach by minimizing the deviation
 in an average sense.
 The problem can then be reduced to a simple quadratic minimization.
 The advantage of this approach is its flexibility in choosing weight
 factors and regions of minimization.
 In addition, it is easily applicable to  more general cases including
 broad classes of functions and also two-step approximations needed in
 the next section.
 In fact, the function $1/x$ refers to $N_f=2$ degenerate quark
 flavours.
 In the general case one has to approximate the function $x^{-N_f/2}$.
 Since in the path integrals for one species of Majorana fermions
 the square root of the fermion determinant appears (see previous
 section), a Majorana fermion is obtained for $N_f=1/2$, hence we
 have to approximate $x^{-1/4}$.
 This corresponds to writing
\be \label{eq23}
|\sqrt{\det(Q)}| = \{\det(Q^\dagger Q)\}^{1/4} =
\{\det(\tilde{Q}^2)\}^{1/4} \ .
\ee
 Here $\tilde Q$ denotes the Hermitean fermion matrix defined by
\be \label{eq24}
\tilde{Q} \equiv \gamma_5 Q = \tilde{Q}^\dagger \ . 
\ee
 (See eq.\ (\ref{eq22}).)

 In what follows, we shall always assume that in the numerical
 simulations the spectrum of $Q^\dagger Q$ is bounded from below, and
 is contained in the interval $[\epsilon,\lambda]$.
 In principle, this may induce some restrictions on the bare
 parameters $\beta,K$, but in practical numerical simulations,
 e.~g.\ in QCD, this condition seems always to be satisfied.
 Before discussing in section \ref{sec4} the ways of implementing the
 fermion determinant (\ref{eq23}) in the path integral, let us
 introduce some classes of the necessary optimized polynomial
 approximations.

\subsection{Single step approximation}             \label{subsec31}
 In order to define an optimized polynomial approximation $P(x)$ of
 the function $x^{-\alpha} \equiv x^{-N_f/2}$ in the above interval,
 the first thing to do is to define a positive norm $\delta$
 characterizing the deviation of two functions from each other.
 The approximation is optimal if $\delta$ is minimal.
 A simple possibility would be to minimize, for instance,
\be \label{eq25}
\int_\epsilon^\lambda dx x^w \left[ x^{-\alpha} - P(x) \right]^2 \ .
\ee
 Here $x^w$ is some arbitrary power giving different weights to
 different parts of the interval.

 In what follows, we shall only consider the special case $w=2\alpha$,
 which corresponds to considering the relative deviation.
 This gives the {\em deviation norm}
\be \label{eq26}
\delta \equiv \left\{
\int_\epsilon^\lambda dx \left[ 1 - x^\alpha P(x) \right]^2
\right\}^\half \ .
\ee
 The square root in the definition of the norm is a matter of
 convention.
 One can, as well, minimize
$$
\delta^2 = 
\int_\epsilon^\lambda dx \left[ 1 - x^\alpha P(x) \right]^2
\equiv \int_\epsilon^\lambda dx \left[ 1 - \sum_{\nu=0}^n 
c_\nu x^{\alpha+n-\nu}  \right]^2 
$$
\be \label{eq27}
= (\lambda-\epsilon) - 2\sum_{\nu=0}^n c_\nu
\frac{\lambda^{1+\alpha+n-\nu}-\epsilon^{1+\alpha+n-\nu}}
{1+\alpha+n-\nu} + 
\sum_{\nu_1=0}^n \sum_{\nu_2=0}^n c_{\nu_1}c_{\nu_2}
\frac{\lambda^{1+2\alpha+2n-\nu_1-\nu_2} -
     \epsilon^{1+2\alpha+2n-\nu_1-\nu_2}}
              {1+2\alpha+2n-\nu_1-\nu_2}  \ .
\ee
 The coefficients of the polynomial $c_\nu$ are assumed to be real.

 The second form in (\ref{eq27}) shows the advantage of choosing norms
 of the type (\ref{eq25})-(\ref{eq26}): since $\delta^2$ is a
 quadratic form of the coefficients, the minimum can be easily
 determined to be at
\be \label{eq28}
c_\nu \equiv c_{n\nu}(\alpha;\epsilon,\lambda) =
\sum_{\nu_1=0}^n M_{(n)\nu\nu_1}^{-1} V_{(n)\nu_1} \ , 
\ee
 where from (\ref{eq27}) we have
$$
V_{(n)\nu} = 
\frac{\lambda^{1+\alpha+n-\nu}-\epsilon^{1+\alpha+n-\nu}}
              {1+\alpha+n-\nu} \ ,
$$
\be \label{eq29} 
M_{(n)\nu_2\nu_1} = M_{(n)\nu_1\nu_2} =
\frac{\lambda^{1+2\alpha+2n-\nu_1-\nu_2} -
     \epsilon^{1+2\alpha+2n-\nu_1-\nu_2}}
              {1+2\alpha+2n-\nu_1-\nu_2}  \ .
\ee
 The optimized polynomial approximation is then
\be \label{eq30}
P_n(\alpha;\epsilon,\lambda;x) \equiv
\sum_{\nu=0}^n c_{n\nu}(\alpha;\epsilon,\lambda) x^{n-\nu} \ .
\ee
 The roots of this polynomial will be needed in the form
\be \label{eq31}
r_{nj}(\alpha;\epsilon,\lambda) \equiv r_j \equiv
(\mu_j + i\nu_j)^2 = \mu_j^2-\nu_j^2 +2i\mu_j\nu_j
\hspace{2em} (j=1,\ldots,n)  \ ,
\ee
 where $\nu_j > 0$ will be assumed.
 Since the coefficients $c_\nu$ are real, the roots come in
 pairs with equal $\nu_j$ and opposite $\mu_j$.
 One can then write
\be \label{eq32}
P_n(\alpha;\epsilon,\lambda;x) =
c_{n0}(\alpha;\epsilon,\lambda) \prod_{j=1}^n 
[x-r_{nj}(\alpha;\epsilon,\lambda)] =
c_{n0} \prod_{j=1}^n[(\sqrt{x}+\mu_j)^2+\nu_j^2] \ . 
\ee

 For a good approximation usually quite high orders $n \gg 1$ are
 necessary.
 In this case the numerical determination of the coefficients
 $c_{n\nu}(\alpha;\epsilon,\lambda)$ and roots
 $r_{nj}(\alpha;\epsilon,\lambda)$ can become cumbersome.
 In particular, very high numerical precision is necessary, which
 is not available e.~g.\ in Fortran.
 However, algebraic programs usually provide the possibility to
 calculate with arbitrarily high precision.
 For instance, using this feature of Maple, one can obtain the
 required precision, and use the results as input data for Fortran
 programs.
 For the calculation of the inverse matrix in (\ref{eq28}) one can
 use LU-decomposition with pivoting \cite{LUDEC}.
 The roots of the polynomial can be numerically determined by
 Laguerre iteration \cite{LAGUERRE}.

\begin{figure}[ht]
\epsfig{file=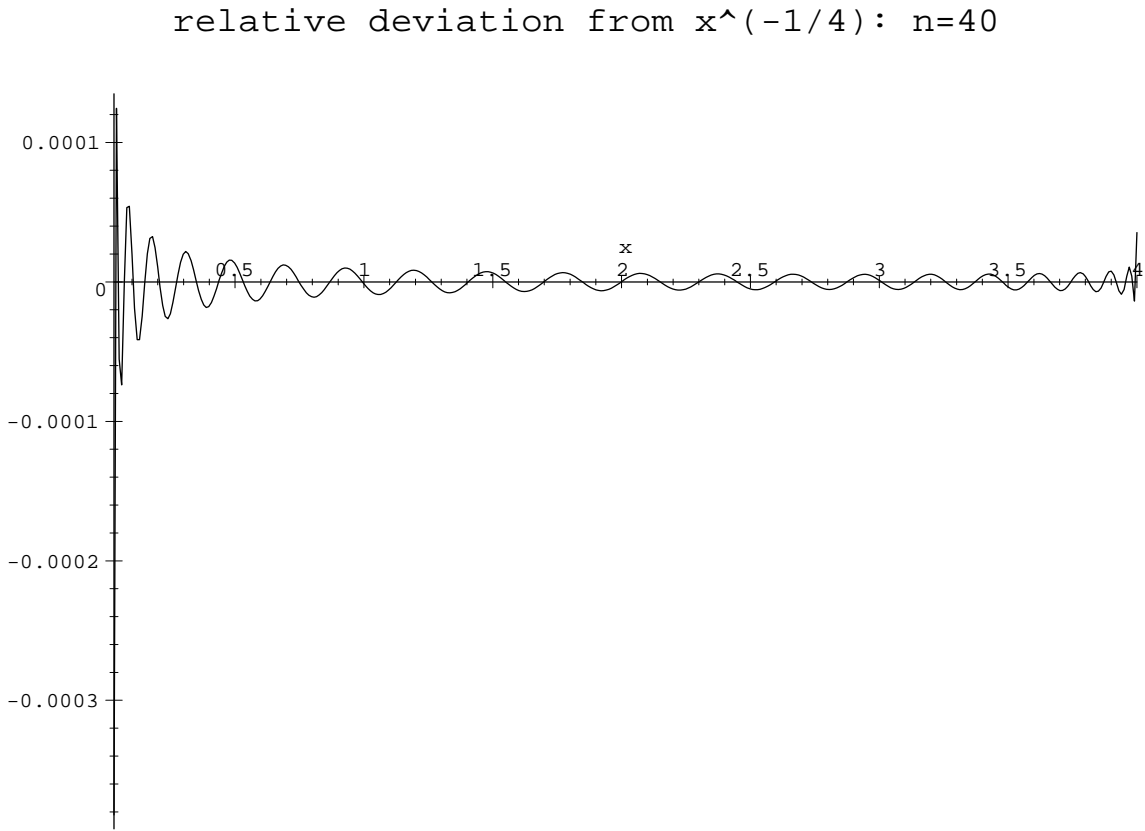,
        bbllx=0,bblly=0,bburx=530,bbury=720,
        width=12.0cm,height=17.0cm,angle=270}
\begin{center}                               
\parbox{16cm}{\caption{ \label{fig1}         
 The relative deviation of $P_{40}(\frac{1}{4};0.03,4.0;x)$ from
 $x^{-\frac{1}{4}}$.
 The value of the deviation norm is here:
 $\delta=2.51.. \cdot 10^{-5}$.
}}            
\end{center}  
\end{figure}

 For gluinos on the lattice we need the power $\alpha=1/4$.
 In this case good approximations can be achieved by polynomials
 of order $n=32$ or $n=40$.
 An example for the spectral interval $[\epsilon=0.03,\lambda=4.0]$,
 which is typical in the test runs discussed in section
 \ref{subsec43}, is illustrated in figure \ref{fig1}.

\subsection{Two-step approximations}               \label{subsec32}
 In the next section two-step approximations will be used for the
 local bosonic description of the fermion determinant.
 Let us assume that a first polynomial approximation $\bar{P}(x)$
 of the function $x^{-\alpha}$ is already known.
 This can be obtained, for instance, by the method described in the
 previous subsection for some relatively low order $\bar{n}$.
 The task is now to find the best approximation in the product form
 $\bar{P}(x) P(x)$, with a polynomial $P(x)$ of order $n > \bar{n}$.

 This can be done, in fact, quite similarly as in subsection
 \ref{subsec31}.
 The equation replacing eq.\ (\ref{eq27}) is now
\be \label{eq33}
\delta^2 = \int_\epsilon^\lambda dx \left[ 1 - \sum_{\nu=0}^n 
\sum_{\bar{\nu}=0}^{\bar{n}} 
c_\nu \bar{c}_{\bar{\nu}} x^{\alpha+n+\bar{n}-\nu-\bar{\nu}}
\right]^2 \ .
\ee
 Here $\bar{c}_{\bar{\nu}}$ denotes the coefficients in $\bar{P}$.
 Everything remains the same as in subsection \ref{subsec31}, only
 (\ref{eq29}) has to be replaced by
$$
V_{(\bar{n},n)\nu} = \sum_{\bar{\nu}=0}^{\bar{n}} \bar{c}_{\bar{\nu}}
\frac{\lambda^{1+\alpha+n+\bar{n}-\nu-\bar{\nu}} -
     \epsilon^{1+\alpha+n+\bar{n}-\nu-\bar{\nu}}}
              {1+\alpha+n+\bar{n}-\nu-\bar{\nu}} \ ,
$$
\be \label{eq34} 
M_{(\bar{n},n)\nu_2\nu_1} = 
\sum_{\bar{\nu_1}=0}^{\bar{n}} \sum_{\bar{\nu_2}=0}^{\bar{n}} 
\bar{c}_{\bar{\nu_1}} \bar{c}_{\bar{\nu_2}} \frac{
 \lambda^{1+2\alpha+2n+2\bar{n}-\nu_1-\nu_2-\bar{\nu_1}-\bar{\nu_2}}-
\epsilon^{1+2\alpha+2n+2\bar{n}-\nu_1-\nu_2-\bar{\nu_1}-\bar{\nu_2}}}
         {1+2\alpha+2n+2\bar{n}-\nu_1-\nu_2-\bar{\nu_1}-\bar{\nu_2}}
 \ .
\ee
 The coefficients of the optimized polynomial 
 $P(x) \equiv P_{\bar{n},n}(\alpha;\epsilon,\lambda;x)$ are
 given by
\be \label{eq35}
c_{\bar{n},n\nu}(\alpha;\epsilon,\lambda) =
\sum_{\nu_1=0}^n M_{(\bar{n},n)\nu\nu_1}^{-1} V_{(\bar{n},n)\nu_1} \ . 
\ee

 A somewhat different type of approximation will also be needed
 when, for the given polynomial $\bar{P}(x)$, the optimized $P(x)$ is
 searched which is minimizing
\be \label{eq36}
\delta^2 = \int_\epsilon^\lambda dx x^w 
\left[ x^\alpha \bar{P}(x)- P(x) \right]^2 \ .
\ee
 In this case, instead of (\ref{eq34}), we have
$$
V_{(\bar{n},n)\nu} = \sum_{\bar{\nu}=0}^{\bar{n}} \bar{c}_{\bar{\nu}}
\frac{\lambda^{1+w+\alpha+n+\bar{n}-\nu-\bar{\nu}} -
     \epsilon^{1+w+\alpha+n+\bar{n}-\nu-\bar{\nu}}}
              {1+w+\alpha+n+\bar{n}-\nu-\bar{\nu}} \ ,
$$
\be \label{eq37} 
M_{(\bar{n},n)\nu_2\nu_1} = 
\frac{\lambda^{1+w+2n-\nu_1-\nu_2}-\epsilon^{1+w+2n-\nu_1-\nu_2}}
              {1+w+2n-\nu_1-\nu_2} \ .
\ee
%

\subsection{Approximations in the complex plane}   \label{subsec33}
 The above polynomial approximation scheme can also be extended to
 the complex plane.
 Denoting the complex variable by $(x+iy)$, it is possible to
 approximate the function $(x+iy)^{-\alpha}$ in some region of the
 complex plane.
 This problem occurs, for instance, in the non-hermitian variants of
 local bosonic algorithms \cite{BORFOR}.
 The approximation regions have to cover the complex spectrum of the
 fermion matrix $Q$.
 Since, according to the relation (\ref{eq22}), the eigenvalues of $Q$
 come in complex conjugate pairs, the region of approximation is
 symmetric with respect to a reflection on the real axis.
 Natural shapes of the complex region are circle or ellipse, but a
 simple symmetric rectangle is also appropriate.
 
 In this latter case, for instance, one can minimize
\be \label{eq38}
\int_\epsilon^\lambda dx \int_{-\gamma}^\gamma dy
|x+iy|^w \left| (x+iy)^{-\alpha} - P(x+iy) \right|^2 \ .
\ee
 Therefore, for $w=2\alpha$, eq.\ (\ref{eq27}) is replaced by
\be \label{eq39}
\delta^2 = \int_\epsilon^\lambda dx \int_{-\gamma}^\gamma dy
\left| 1 - \sum_{\nu=0}^n 
c_\nu (x+iy)^{\alpha+n-\nu} \right|^2  \ . 
\ee
 In the interesting cases, when $\alpha=N_f=\half,1,2,\ldots$, the
 integrals can be performed analytically and give rise to relatively
 simple expressions in $\lambda,\epsilon,\gamma$.
 The coefficients $c_{n\nu}(\alpha;\epsilon,\lambda,\gamma)$ of
 the optimal polynomial approximation
 $P_n(\alpha;\epsilon,\lambda,\gamma;x+iy)$ are given again by
 a similar expression as eq.\ (\ref{eq28}).

\begin{figure}[ht]
\epsfig{file=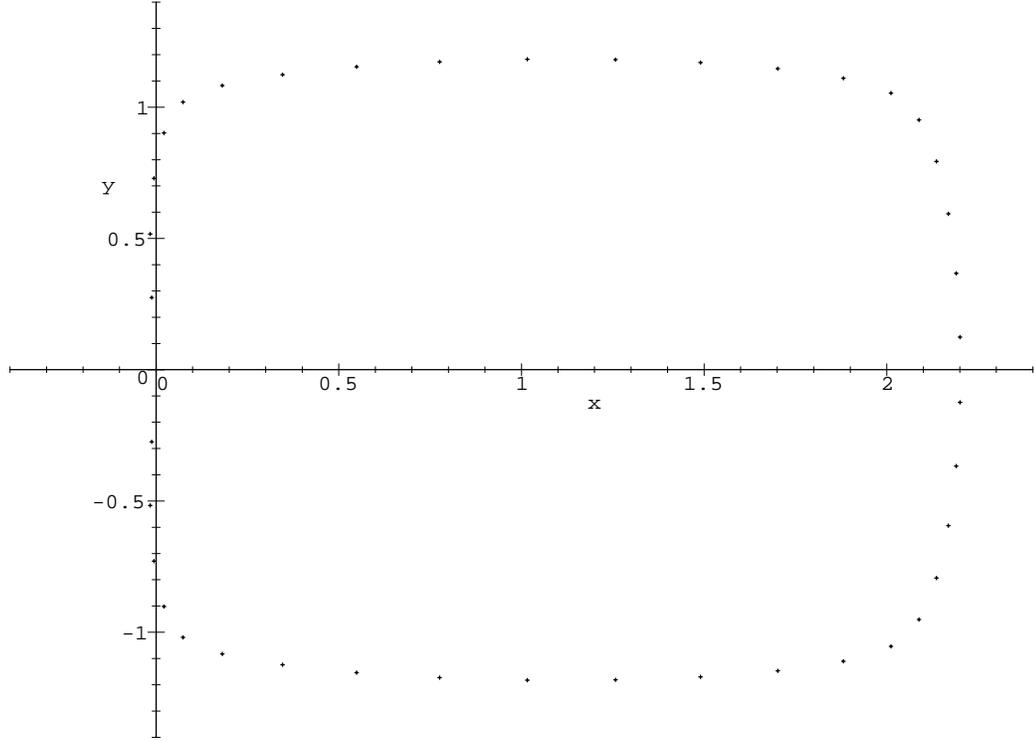,
        bbllx=0,bblly=0,bburx=530,bbury=720,
        width=12.0cm,height=17.0cm,angle=270}
\begin{center}                               
\parbox{16cm}{\caption{ \label{fig2}         
 The positions of the roots of $P_{40}(1;0.1,2.0,1.0;x+iy)$
 in the complex plane.
}}            
\end{center}  
\end{figure}

 An illustration for the positions of the roots of the polynomial
 $P_n$ for $\alpha=1$, $(\epsilon=0.1,\; \lambda=2.0,\; \gamma=1.0)$
 is figure \ref{fig2}.
 The spectral region is typical for the situation in the tests
 discussed in section \ref{subsec43}, similarly to fig.\ \ref{fig1}.
 As one can see, the roots wind around the rectangular region
 in the complex plane.
 The value of the deviation norm is $\delta=0.0132..$.
 This is relatively large for this high order, showing that the
 rectangular region is not optimal for this kind of approximations.
 Elliptical shapes are more natural from the mathematical point
 of view.
 A rectangle is, however, more convenient from the point of view
 of monitoring the eigenvalue ranges.
 For instance, one can easily determine the extremal eigenvalues 
 of the Hermitean matrices $(Q+Q^\dagger)$ and $i(Q^\dagger-Q)$,
 and infer from them the necessary rectangle.
 (The extremal eigenvalues can be determined, for instance, by
 gradient methods \cite{FADEEV}.)
 In addition, as a closer inspection shows, most of the
 contribution to $\delta$ comes from the areas of the rectangle
 near the corners, where there are typically no eigenvalues anyway.
 In fact, in the important central regions the approximation is
 several orders of magnitude better.
                                                                     
\section{Local bosonic algorithms}                     \label{sec4}
 Let us now turn to the question, how to implement the fermion
 determinant factor in eq.\ (\ref{eq23}) in the path integrals
 appearing, for instance, in the expectation values
 (\ref{eq18})-(\ref{eq21}).
 Following ref.\ \cite{LUSCHER} and taking some polynomial
 approximations, as the Chebyshev polynomials or the polynomials
 introduced in section \ref{subsec31}, we have
\be \label{eq40}
\{ \det \tilde{Q}^2 \}^{1/4} \;\stackrel{\delta}{\longrightarrow}\;
\frac{1}{\det[P_n(\tilde{Q}^2)]} =
\frac{1}{\det(c_{n0})\; \prod_{j=1}^n
\det[(\tilde{Q}+\mu_j)^2+\nu_j^2]}  \ .
\ee
 Here $\;\stackrel{\delta}{\longrightarrow}\;$ means a polynomial
 approximation with deviation norm equal to $\delta$.
 The determinant factors can be written with the help of complex
 {\em pseudofermion fields} $\phi_{jx},\; (j=1,..,n)$ as Gaussian
 integrals:
\be \label{eq41}
\frac{1}{\det[(\tilde{Q}+\mu_j)^2+\nu_j^2]} \propto
\int [d\phi_j] \exp\left\{ -\sum_{xy} \phi_{jy}^*
[\nu_j^2\delta_{yx} + (\tilde{Q}+\mu_j)_{yx}^2] \phi_{jx} 
\right\}  \ .
\ee
 This is how the fermion determinant is represented in path
 integrals by a local bosonic action.

\subsection{Two-step approximation with noisy correction}
\label{subsec41}
 In order to obtain a very high precision approximation one has to
 take a large number of pseudofermion fields.
 This increases the required storage space and might also cause
 problems with long autocorrelations. 
 The solution is to introduce a two-step approximation scheme, as
 discussed in section \ref{subsec32}.

 In the first step we make an approximation by the polynomial
 $\bar{P}(x)$ with a deviation norm $\bar{\delta}$:
\be \label{eq42}
\{ \det \tilde{Q}^2 \}^{1/4} 
\;\stackrel{\bar{\delta}}{\longrightarrow}\;
\frac{1}{\det[\bar{P}(\tilde{Q}^2)]}  \ .
\ee
 This is improved to a smaller deviation norm $\delta$ in the second
 step by multiplying with the polynomial $P(x)$:
\be \label{eq43}
\{ \det \tilde{Q}^2 \}^{1/4} \;\stackrel{\delta}{\longrightarrow}\;
\frac{1}{\det[\bar{P}(\tilde{Q}^2)] \det[P(\tilde{Q}^2)]} \ .
\ee
 The first approximation is realized by pseudofermion fields as in
 eq.\ (\ref{eq41}), whereas the correction factor
 $\det[P(\tilde{Q}^2)]$ is taken into account by a Metropolis
 correction step, similarly to ref.\ \cite{KENKUT}.
 Since this latter is realized by Gaussian i.~e.\ ``noisy''
 {\em unbiased estimators}, we shall call it here briefly
 {\em noisy correction}.

 By the Monte Carlo updating process one has to reproduce the
 canonical gauge field distribution
\be \label{eq44}
w[U] = \frac{e^{-S_g[U]}}
{\det[\bar{P}(\tilde{Q}[U]^2)] \det[P(\tilde{Q}[U]^2)]}  \ ,
\ee
 where $S_g[U]$ is the pure gauge field part of the action in
 (\ref{eq04}).
 Using (\ref{eq41}) to represent $\det\bar{P}$, the canonical
 distribution in terms of the gauge field $U$ and pseudofermion
 field $\phi$ can be written in the form
\be \label{eq45}
w[U,\phi] = \frac{e^{-S_{g\bar{P}}[U,\phi]}}
{\det[P(\tilde{Q}[U]^2)]}  \ .
\ee
 The updating of the $\phi$ field can be straightforwardly done
 by heatbath and/or overrelaxation algorithms.

 In order to prove {\em detailed balance} for the updating of the
 gauge field, we have to show that the transition probability
 $P([U^\prime] \lar [U])$ of the Markov process satisfies
\be \label{eq46}
\frac{P([U] \lar [U^\prime])}{P([U^\prime] \lar [U])} =
\frac{e^{-S_{g\bar{P}}[U]} \det[P(\tilde{Q}[U^\prime]^2)]}
{e^{-S_{g\bar{P}}[U^\prime]} \det[P(\tilde{Q}[U]^2)]}  \ .
\ee
 The transition probability is given by
\be \label{eq47}
P([U^\prime] \lar [U]) = 
P_M([U^\prime] \lar [U]) P_A([U^\prime] \lar [U]) \ ,
\ee
 where $P_M([U^\prime] \lar [U])$ is the transition probability
 of the updating of the gauge field with the action
 $S_{g\bar{P}}[U]$, and $P_A([U^\prime] \lar [U])$ is the acceptance
 probability of the correction step.
 $P_M([U^\prime] \lar [U])$ is realized by standard Metropolis
 sweeps.
 It is assumed to satisfy detailed balance for the action
 $S_{g\bar{P}}[U]$, that is
\be \label{eq48}
\frac{P_M([U] \lar [U^\prime])}{P_M([U^\prime] \lar [U])} =
\frac{e^{-S_{g\bar{P}}[U]}}{e^{-S_{g\bar{P}}[U^\prime]}}  \ .
\ee
 This can be achieved in different ways, for instance, by a randomly
 chosen order of links or by updating always one half of the
 links in a checkboard decomposition or by doing the sweeps pairwise
 with opposite orders of link updates.
 Comparing eqs.\ (\ref{eq46})-(\ref{eq48}) one can see that the
 condition for the acceptance probability is
\be \label{eq49}
\frac{P_A([U^\prime] \lar [U])}{P_A([U] \lar [U^\prime])} =
\frac{\det[P(\tilde{Q}[U]^2)]}{\det[P(\tilde{Q}[U^\prime]^2)]} \ .
\ee
 This could be satisfied by the the standard Metropolis acceptance
 probability
\be \label{eq50}
P_A([U^\prime] \lar [U]) = \min \left\{ 1,
\frac{\det[P(\tilde{Q}[U]^2)]}{\det[P(\tilde{Q}[U^\prime]^2)]}
\right\}  \ .
\ee
 However, this would be too expensive to implement numerically.

 The idea of the noisy correction \cite{KENKUT} is to generate a
 random vector $\eta$ according to the normalized Gaussian
 distribution
\be \label{eq51}
\frac{e^{-\eta^\dagger P(\tilde{Q}[U]^2)\eta}}
{\int [d\eta] e^{-\eta^\dagger P(\tilde{Q}[U]^2)\eta}}  \ ,
\ee
 and to accept the change $[U^\prime] \lar [U]$ with probability
\be \label{eq52}
\min\left\{ 1,A(\eta;[U^\prime] \lar [U]) \right\} \ ,
\ee
 where
\be \label{eq53}
A(\eta;[U^\prime] \lar [U]) =
\exp\left(-\eta^\dagger\{ P(\tilde{Q}[U^\prime]^2) -
                           P(\tilde{Q}[U]^2) \}\eta\right)  \ .
\ee
 This means that the acceptance probability is
\be \label{eq54}
P_A([U^\prime] \lar [U]) = 
\frac{\int_{A>1} [d\eta] e^{-\eta^\dagger P(\tilde{Q}[U]^2)\eta} +
      \int_{A<1} [d\eta] e^{-\eta^\dagger 
P(\tilde{Q}[U^\prime]^2)\eta}}
{\int [d\eta] e^{-\eta^\dagger P(\tilde{Q}[U]^2)\eta}}  \ .
\ee
 Due to the relation
\be \label{eq55}
A(\eta;[U] \lar [U^\prime]) = \frac{1}{A(\eta;[U^\prime] \lar [U])}
\ee
 the detailed balance condition in (\ref{eq49}) is satisfied.

 In this form the algorithm would be still too expensive, because of
 the necessity to generate the distribution in eq.\ (\ref{eq51}).
 One can, however, easily generate the simple Gaussian distribution
\be \label{eq56}
\frac{e^{-\eta^{\prime\dagger}\eta^\prime}}
{\int [d\eta^\prime] e^{-\eta^{\prime\dagger}\eta^\prime}}  \ ,
\ee
 and then obtain the above $\eta$ from
\be \label{eq57}
\eta = P(\tilde{Q}[U]^2)^{-\half} \eta^\prime  \ .
\ee
 The exact calculation is still too difficult, but one can obtain
 the result to a required approximation by the two-step polynomial
 approximation technique introduced at the end of section
 \ref{subsec32}.
 An approximation with the same deviation norm $\delta$ as in 
 (\ref{eq43}) is sufficient.

 According to eq.\ (\ref{eq43}) the polynomial approximation $R(x)$
 for $P(x)^{-1/2}$ has to obey
\be \label{eq58}
R(x) \;\stackrel{\delta}{\longrightarrow}\; P(x)^{-\half}
\;\stackrel{\delta}{\longrightarrow}\; 
x^{\frac{1}{8}}\bar{P}(x)^\half 
\;\stackrel{\delta}{\longrightarrow}\; 
x^{\frac{1}{8}}S(\bar{P}(x)) \ . 
\ee
 Here we defined the approximation
\be \label{eq59}
S(\bar{P}) \;\stackrel{\delta}{\longrightarrow}\; 
\bar{P}^\half  \ ,
\ee
 which can be obained by a single-step polynomial approximation
 defined in section \ref{subsec31} for $\alpha=-\half$ in the
 interval $[\lambda^{-1/4},\epsilon^{-1/4}]$.
 After this, $R(x)$ in (\ref{eq58}) can be determined by minimizing
 the deviation norm $\delta$ defined in eq.\ (\ref{eq36}) for
 $\alpha=1/8$ and $\bar{P}(x)$ replaced by $S(\bar{P}(x))$.
 The weight factor is somewhat arbitrary, but a good choice is, for
 instance, $w=0$.
 This means that in the interval $[\epsilon,\lambda]$ the absolute
 deviation is minimized. 

 This algorithm based on two-step approximations is not ``exact''
 because for any finite $\delta$ there is still some deviation from
 the functions required in eqs.~(\ref{eq43}) and
 (\ref{eq58}-\ref{eq59}).
 This is in contrast to the exact algorithms with noisy correction
 proposed for QCD in refs.~\cite{BORFOR,KENKUT}.
 It is, however, not clear how to extend the exact algorithms to
 the case with square roots considered in this paper.
 Nevertheless the two-step approximations can easily be made very
 precise.
 Namely, provided that the second approximation is already good,
 the acceptance rate in the noisy correction step mainly depends
 on the first polynomial in (\ref{eq42}), and only very little
 on the precision of the approximation achieved by the second
 one.
 The application of the polynomials in eq.~(\ref{eq53}) and
 (\ref{eq58}) is not very time consuming, therefore the decrease of
 $\delta$  does not lead to a dangerous slowing down of the
 simulation.
 This applies both to the gluino model considered in this paper
 and to QCD with quarks in the fundamental representation.
 Therefore the two-step algorithm proposed here is also a
 potential candidate for simulating QCD.

\subsection{Noisy estimators}                         \label{subsec42}
 The pseudofermion fields can also be used to obtain the matrix
 elements of the fermion propagators.
 The required expectation values are easily obtained during the 
 updating process, and in this way the fermion matrix elements can be
 determined.
 The expressions in terms of the pseudofermion fields are usually
 called {\em noisy estimators}.

 The starting relation is, for a general matrix $M$, 
\be \label{eq60}
\frac{\partial}{\partial M_{\alpha\beta}} \det M =
M^{-1}_{\beta\alpha} \det M  \ .
\ee
 In our case we have from this
\be \label{eq61}
\tilde{Q}^{-1}_{yx} = \frac{ 2
\frac{\partial}{\partial \tilde{Q}_{xy}} [\det\tilde{Q}^2]^{1/4}}
{[\det\tilde{Q}^2]^{1/4}}  \ .
\ee
 Combining this formula with eqs.\ (\ref{eq40})-(\ref{eq41}) we
 obtain
\be \label{eq62}
\left\langle \tilde{Q}^{-1}_{yx} \right\rangle_U =
\left\langle Q^{-1}_{yx}\gamma_5 \right\rangle_U = 
\left\langle \psi_y \overline{\psi}_x\gamma_5 \right\rangle =
- 2\sum_{j=1}^n \left\langle \overline{\phi}_{jy}\phi^*_{jx}
+ \phi_{jy}\overline{\phi}^*_{jx} \right\rangle_{\phi U}  \ .
\ee
 Here the notation
\be \label{eq63}
\overline{\phi}_{jx} \equiv \sum_y  (\tilde{Q}+\mu_j)_{xy} 
\phi_{jy} 
\ee
 is used.
 For instance, for the gluino condensate eqs.\ (\ref{eq20}) and
 (\ref{eq62}) give
\be \label{eq64}
- \left\langle (\overline{\Psi}_x \Psi_x) \right\rangle =
- 2\sum_{j=1}^n \left\langle (\phi^*_{jx}\gamma_5\overline{\phi}_{jx})
+ (\overline{\phi}^*_{jx}\gamma_5 \phi_{jx}) \right\rangle_{\phi U}
  \ .
\ee
 (Note that this is the {\em unrenormalized} condensate, which has
 to be subtracted and renormalized, in order to obtain a physical
 quantity.)

 The expectation values of several pairs of fermion variables can
 also be obtained in a similar way.
 For instance, for two pairs one can use
\be \label{eq65}
\frac{\partial^2}{\partial M_{\alpha\beta}\partial M_{\gamma\delta}} 
\det M =
\left( M^{-1}_{\beta\alpha}M^{-1}_{\delta\gamma}
     - M^{-1}_{\beta\gamma}M^{-1}_{\delta\alpha} \right) \det M \ ,
\ee
 and in our case
\be \label{eq66}
\frac{4}{3} \left(
\frac{\partial^2}{\partial \tilde{Q}_{yx}\partial \tilde{Q}_{wz}} -
\frac{\partial^2}{\partial \tilde{Q}_{wx}\partial \tilde{Q}_{yz}}
\right) [\det \tilde{Q}^2]^{1/4} =
\left( \tilde{Q}^{-1}_{xy}\tilde{Q}^{-1}_{zw}
     - \tilde{Q}^{-1}_{xw}\tilde{Q}^{-1}_{zy} \right) 
[\det \tilde{Q}^2]^{1/4}  \ .    
\ee
 As in (\ref{eq62}), it is straightforward to transform this in an
 expectation value in terms of the pseudofermion field.
%
\begin{table}[ht]
\begin{center}
\parbox{14cm}{\caption{ \label{tab1}
 The values of some simple characteristic quantities on $4^3 \cdot 8$
 lattice at $\beta=2.0,\; K=0.150$.
 The quantities are defined in the text.
 In the sixth column the number of combined sweeps is given in
 thousands.
 For the two-step algorithms the seventh column shows the updating
 probability of gauge links, and the last column contains the
 acceptance rate of the noisy correction step.
}}
\end{center}
\begin{center}
\begin{tabular}{|c|l|l|l|l|c|c|c|}
\hline
  $(\bar{n},)n$  &  $\hspace{2em}\Box$  &  $\hspace{1.5em}|P_l|$  &
  $\hspace{0.5em}-(\overline{\Psi}_x \Psi_x)$  &
  $\hspace{2em}\chi_2$  &  sweep  &  $p_u$  &  acc.  \\
\hline\hline
 2  & 0.50068(19) & 0.05088(23) & 9.4887(24) & 0.54831(93) &
 28 &  &  \\
\hline
 4  & 0.51117(16) & 0.05095(23) & 11.1486(45) & 0.5668(10) &
 40 &  &  \\
\hline
 8  & 0.50857(38) & 0.05191(51) & 11.818(12)  & 0.5698(24) &
 20 &  &  \\
\hline
 16 & 0.50675(8)  & 0.0488(8)   & 11.908(24)  & 0.5657(39) &
 20 &  &  \\
\hline
 24 & 0.5075(14)  & 0.0524(16)  & 11.897(45)  & 0.5619(72) &
 15  &  &  \\ 
\hline\hline
 4,24 & 0.50611(29)  & 0.0506(3) & 11.160(4)  & 0.5561(13) &
 60  &  0.1  &  0.59  \\
\hline
 6,24 & 0.50578(32)  & 0.0514(3) & 11.639(8)  & 0.5563(14) &
 25  &  1.0  &  0.54  \\
\hline
 8,24 & 0.50600(40)  & 0.0508(6) & 11.833(12) & 0.5556(21) &
 20  &  1.0  &  0.74  \\ 
\hline\hline
\end{tabular}
\end{center}
\end{table}

\subsection{Numerical simulation tests}               \label{subsec43}
 The first tests of the algorithm described in the previous
 subsections were performed in the SU(2) Yang Mills model with
 gluinos.
 The bare parameters were $\beta=2.0$ and $0.125 \leq K \leq 0.175$.
 The lattice sizes were in the range $4^3 \cdot 8$ to $8^3 \cdot 16$.

 In order to observe the $n$-dependence of some simple global
 quantities and the corresponding autocorrelations, a series of tests
 were done on $4^3 \cdot 8$ lattice at $(\beta=2.0,K=0.150)$.
 This is in a reasonable range of parameters for numerical
 simulations, as the obtained estimates of the value of the string
 tension in lattice units show (see below).
 For the spectral interval $[\epsilon=0.03,\; \lambda=4.0]$ was
 taken, which contains the eigenvalues of $\tilde{Q}^2$ confortably:
 it turned out that in equilibrium configurations the eigenvalues
 were always larger than 0.05 and smaller than 3.7 (they were
 monitored at every sweep by a gradient algorithm \cite{FADEEV}). 
 The chosen quantities were: the plaquette
 $\Box \equiv W_{1,1} \equiv \half{\rm Tr\,}U_{pl}$, the absolute
 value of the Polyakov line in the time ($L_t=8$) direction
 $|P_l|$, the unrenormalized gluino condensate from the noisy
 estimator in eq.\ (\ref{eq64}), and the $k \otimes k$ Creutz ratio
 with $k=2$ calculated from the Wilson loops as
\be \label{eq67}
\chi_k \equiv \frac{W_{k,k}W_{k-1,k-1}}{W_{k,k-1}W_{k-1,k}} \simeq
e^{-a^2\sigma} \ .
\ee
 Here $a^2\sigma$ is the string tension in lattice units.

 The results with $2 \leq n \leq 24$ for the single-step algorithm,
 and with $\bar{n}=4,6,8;\; n=24$ for the two-step algorithm
 discussed in subsection \ref{subsec41} are shown in table \ref{tab1}.
 The corresponding integrated autocorrelations are given in table
 \ref{tab2}.
\begin{table}[ht]
\begin{center}
\parbox{14cm}{\caption{ \label{tab2}
 The integrated autocorrelations of different quantities expressed
 in numbers of combined sweeps on $4^3 \cdot 8$ lattice at
 $\beta=2.0,\; K=0.150$.
 The quantities are defined in the text.
}}
\end{center}
\begin{center}
\begin{tabular}{|c|c|c|c|c|}
\hline
  $(\bar{n},)n$  &  $\Box$  &  $|P_l|$  &
  $-(\overline{\Psi}_x \Psi_x)$  &  $\chi_2$  \\
\hline\hline
 2  & 6.4(3) & 1.0(1) & 1.0(1) & 2.8(1) \\
\hline
 4  & 13.9(3) & 1.4(1) & 1.5(1) & 5.2(1) \\
\hline
 8  & 36(2) & 3.3(3) & 1.6(1) & 15(1)  \\
\hline
 16 & 117(14) & 8.3(3) & 1.9(1) & 35(4)  \\
\hline
 24 & 165(35) & 19(2)  & 2.3(2) & 43(5)  \\
\hline\hline
 4,24 & 29(3) & 4.0(3) & 1.6(3) & 12.8(6)  \\
\hline
 6,24 & 18(2) & 1.7(4) & 1.8(3) & 7.8(7)  \\
\hline
 8,24 & 23(4) & 2.7(5) & 1.7(4) & 9.1(6) \\
\hline\hline
\end{tabular}
\end{center}
\end{table}

 No particular effort was made to optimize the mixture of the
 different updating steps.
 This should certainly be an important part of the final optimization
 \cite{BJJLSS,BORFOR}, which should be done in larger scale
 applications.
 Here only the $n$-dependence was investigated for a fixed reasonably
 looking mixture: 1 heatbath and 6 overrelaxation sweeps for the
 pseudofermions followed by 2 Metropolis sweeps with 8 hits per link
 for the gauge field.
 In case of the two-step algorithm before the accept-reject correction
 step 2 Metropolis sweeps were done with opposite orders of links.
 This was repeated 5-times. 
 In order to reach a good acceptance, at every link it was first
 decided whether to update it or not.
 By choosing the updating probability $p_u \leq 1$, the acceptance
 rate could be tuned appropriately.
 In table \ref{tab1} the numbers of ``combined sweeps'' always mean
 the numbers of these sequences of sweeps.

 Tables \ref{tab1} and \ref{tab2} show that for the given statistical
 errors $n=24$ is enough, and from the point of view of
 autocorrelations the algorithm with two-step approximation and noisy
 correction is better.
 In addition, compared to a single-step algorithm with the same
 approximation order $n$, the combined sweeps of the two-step
 algorithm also need less computer time.
 In table \ref{tab2} it is amazing to observe that the physically
 more interesting Creutz ratio $\chi_2$ has definitely smaller
 integrated autocorrelations than the plaquette $\Box$.
 The noisy estimators for $-(\overline{\Psi}_x\Psi_x)$ always have
 very short autocorrelations.

 In the two-step algorithm it is an interesting question, how the
 acceptance probability of the noisy correction is behaving for
 increasing lattice sizes.
 In order to see this, the simulation with $\bar{n}=4,\; n=24$
 was scaled up to $6^3 \cdot 12$ and $8^3 \cdot 16$, always with
 $n=24$ in the correction step.
 It turned out that acceptances similar to $\bar{n}=4$ on
 $4^3 \cdot 8$ could be reached with $\bar{n}=6$ and $\bar{n}=8$ on
 $6^3 \cdot 12$ and $8^3 \cdot 16$, respectively.
 With the same mixture of sweeps and link updating probability
 $p_u=0.1$, as in table \ref{tab1}, the acceptance rates were 55\%
 and 65\%, respectively.
 This is good news, which tells that the number of pseudofermion
 fields for the first approximation has to grow at most linearly
 with the lattice extension.

 Another important question is the behaviour of these local bosonic
 algorithms for decreasing lattice spacing and/or fermion mass.
 A detailed investigation of this could not be carried out here.
 From the data in table \ref{tab1} follows that the square root of
 the string tension in lattice units is at $(\beta=2.0,K=0.150)$
 roughly $a\sqrt{\sigma} \simeq 0.75$.
 Defining the physical scale by $\sqrt{\sigma} \simeq 0.45\; GeV$, we
 obtain $a \simeq 1.7\; GeV^{-1}$.
 In order to see the behaviour of the algorithm and of
 $a\sqrt{\sigma}$ as a function of the fermion mass ($K$), numerical
 simulations on $6^3 \cdot 12$ lattices were performed at $\beta=2.0$
 and $K=0.125,\; 0.150,\; 0.175$ with the single-step algorithm.
 The approximations were done by the polynomials
 $P_8(1/4;\epsilon=0.07,\lambda=3.5;x)$,
 $P_8(1/4;\epsilon=0.03,\lambda=4.0;x)$ and
 $P_8(1/4;\epsilon=0.005,\lambda=5.0;x)$, respectively.
 In all three cases the chosen spectral intervals comfortably covered
 the observed extrema of the spectrum of $\tilde{Q}^2$.
 About 1000 combined sweeps were done after equilibration.
 The plaquette autocorrelations were about 10, 30 and 50 complete
 sweeps, respectively.
 At $K=0.125$ and $K=0.150$ the square root of the string tension in
 lattice units were roughly the same as the above value.
 At $K=0.175$ a small decrease to a value $a\sqrt{\sigma} \simeq 0.7$
 was observed.

 Another question is the size of the effects of dynamical gluinos
 on the expectation values.
 In the considered range of bare parameters this is still small.
 A short simulation in pure gauge theory gives, for instance, at
 $\beta=2.0$ on $8^3 \cdot 16$ lattice $\Box = 0.50113(6)$ and
 $\chi_2=0.5490(2)$.
 These are only slightly below the values in table~\ref{tab1}.
 Larger effects are expected at larger $\beta$ in the critical
 region of the hopping parameter $K$. 

 The experience with the local bosonic algoritms on these relatively
 small lattices at relatively low lattice cut-offs was quite positive.
 As remarked above, from the point of view of autocorrelations and
 computational speed the two-step algorithm with noisy correction
 was clearly better.
 Of course, detailed tests on larger lattices and at smaller lattice
 spacings are necessary.
 A direct comparison with the conventional algorithms based on
 discretized classical equations of motion \cite{HCL} would also be
 interesting.
 
\section{Other applications of the optimized polynomials}  \label{sec5}
\subsection{Optimized solvers for matrix inversion}  \label{subsec51}
 The optimized polynomial approximations defined in section \ref{sec3}
 can also be used for matrix inversion instead of the Chebyshev
 polynomials in Chebyshev iteration schemes \cite{VARMAR}.
 In numerical simulations of fermionic field theories an important
 task is to calculate the fermion propagator matrix elements in a
 given gauge field background $Q[U]^{-1} \equiv Q^{-1}$.
 An approximate solution of the equation
\be \label{eq68}
Qp = v
\ee
 is the vector
\be \label{eq69}
p_n = P_n(1;\epsilon,\lambda,\gamma;Q)v =
c_{n0} (1;\epsilon,\lambda,\gamma) \left( \prod_{j=1}^n [Q -
 r_{nj}(1;\epsilon,\lambda,\gamma)] \right) v  \ .
\ee
 Here we used the first form of the optimized polynomial $P_n$ in
 (\ref{eq32}), but other forms are also possible.
 Since we are actually considering the non-Hermitean fermion matrix
 $Q$, the approximation has to be optimized in the region of the
 complex plane $x \in [\epsilon,\lambda],\; y \in [-\gamma,\gamma]$,
 which covers the spectrum of $Q$.
 The deviation norm $\delta$, which gives the precision of the
 approximation in (\ref{eq69}), can be characterized by an expression
 similar to (\ref{eq39}) weighted by the spectral density of the
 matrix $Q$.

\begin{figure}[ht]
\epsfig{file=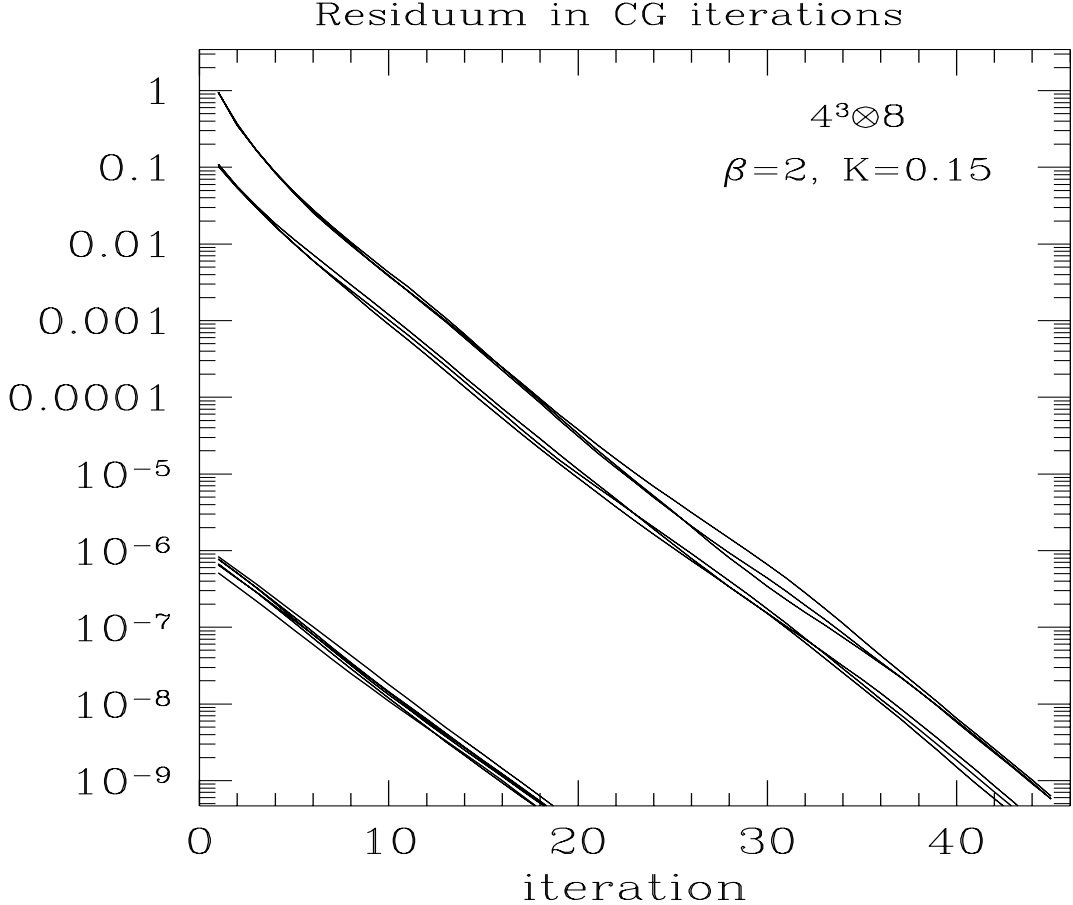,
        width=15.0cm,height=12.0cm,angle=0}
\begin{center}                               
\parbox{16cm}{\caption{ \label{fig3}
 The square of the residuum vector in a CG iteration on a
 $4^3 \cdot 8$ configuration at $(\beta=2.0,K=0.150)$.
 Twelve initial vectors are taken, starting from a randomly chosen
 point.
 The two upper bunches of curves belong to the normal CG iteration,
 whereas in the lower bunch the CG iteration is started after the
 multiplication of the twelve initial vectors by the optimized
 solver $P_{32}(1;0.1,2.0,1.0;x+iy)$ having $\delta=0.024..$.
}}            
\end{center}  
\end{figure}

 For illustrative purposes figure~\ref{fig3} shows the squared
 length of the residue vector
\be \label{eq70}
r_n \equiv v-Qp_n
\ee
 obtained from (\ref{eq69}) on a typical $4^3 \cdot 8$ configuration
 at $(\beta=2.0,K=0.150)$.
 For comparison, the same quantity is also shown as a function of the
 number of iterations in the popular conjugate gradient (CG)
 algorithm.
 As the figure shows, an application of the {\em optimized solver}
 $P_{32}$ reduces the residuum roughly as much as 32 CG iterations.

\begin{figure}[ht]
\epsfig{file=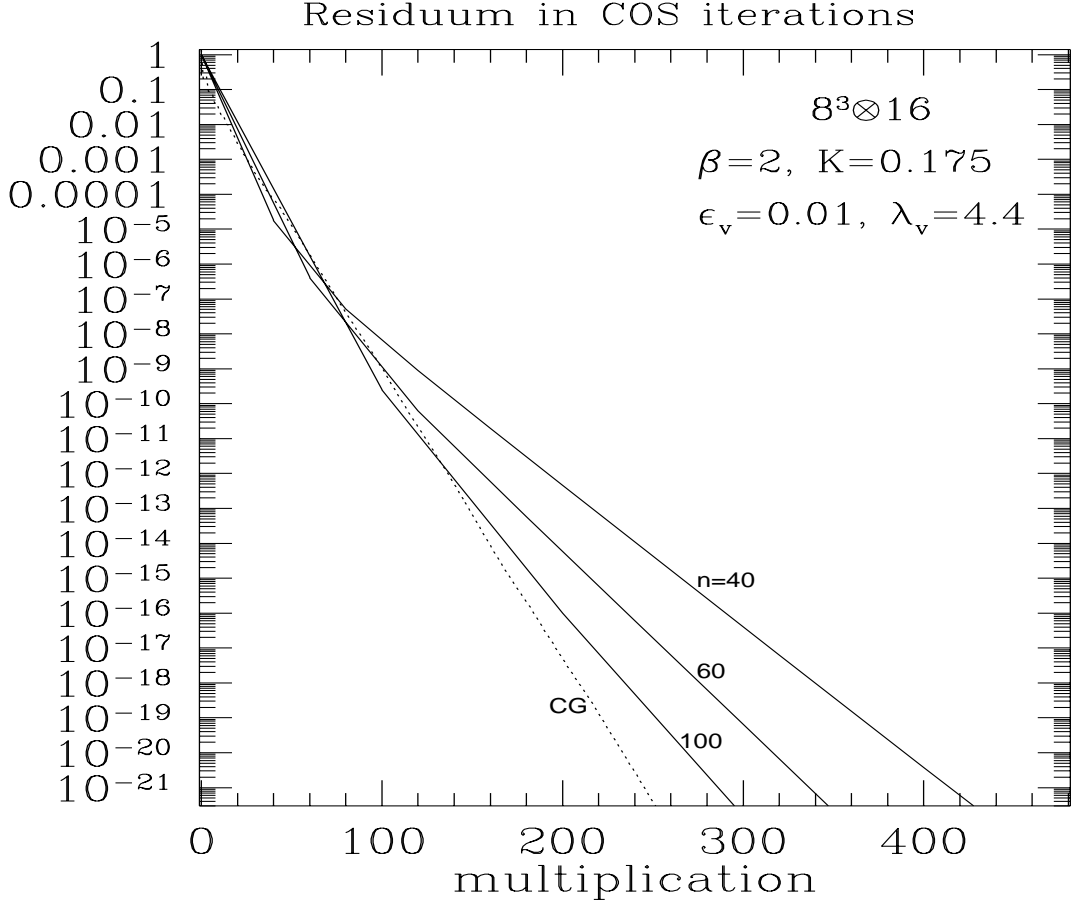,
        width=15.0cm,height=12.0cm,angle=0}
\begin{center}                               
\parbox{16cm}{\caption{ \label{fig4}
 The square of the residuum vector in the cyclic optimized solver
 iteration scheme for different polynomial orders
 $n=40,60,100$.
 The vector $v$ is always the same and the configuration is at
 $\beta=2.0, K=0.175$ on $8^3 \cdot 16$ lattice.
 The interval of polynomial approximations is $[0.01,4.4]$.
 The dotted curve shows the result of conjugate gradient iteration.
}}            
\end{center}  
\end{figure}

 An iterative scheme is obtained if the optimized polynomials are
 successively applied to the residuum vectors.
 The resulting {\em cyclic iteration process} \cite{VARMAR} can be
 represented by the following simple iteration equations:
 start by defining $p_n^{(1)} \equiv p_n$ and $r_n^{(0)} \equiv r_n$
 and then set for $k=1,2,\ldots$ 
$$
r_n^{(k)} = v - Qp_n^{(k)} \ ,
$$
\be \label{eq71}
p_n^{(k+1)} = p_n^{(k)} + P_n r_n^{(k)} \ .
\ee
 $p_n^{(k)}$ tends to the solution $p$ of (\ref{eq68}) in the limit
 $k \to \infty$ because
\be \label{eq72}
p_n^{(k)} = P_n v + (1-P_n Q)p_n^{(k-1)} =
P_n \sum_{j=0}^{k-1} (1-P_n Q)^j v \ .
\ee
 Therefore the iteration is realizing the simple identity
\be \label{eq73}
Q^{-1} = P_n (Q P_n)^{-1} = P_n [ 1-(1-P_n Q)]^{-1} \ .
\ee
 For an illustration of this {\em cyclic optimized solver} (COS)
 iteration scheme see figure~\ref{fig4}, where instead of $Q$ the
 positive matrix $Q^\dagger Q$ is inverted.
 In the figure the square of the residuum vector is shown as a
 function of the number of matrix multiplications (in this case a CG
 iteration step requires only one matrix multiplication).
 As the figure shows, the COS iteration with a high enough order
 polynomial has a similar performance as CG iteration.

 The application of the optimized solver $P_n$ requires $n$
 matrix multiplications and no vector norm calculations.
 Therefore the amount of arithmetics in COS iterations compares
 favourably with the CG iteration.
 Nevertheless there are also known methods which are better than CG
 iteration \cite{BICONJ}, therefore the best choice of the inversion
 method at given bare parameters and lattice sizes has to be decided
 on a case by case basis.
 
 The optimized solver can also be considered as an {\em optimized
 hopping parameter expansion}.
 Let us write the fermion matrix in (\ref{eq09}) as 
\be \label{eq74}
Q \equiv 1 - KM \ ,
\ee
 with the {\em hopping matrix} $M$.
 Then we have
\be \label{eq75}
P_n(1;\epsilon,\lambda,\gamma;Q) =
c_{n0} (1;\epsilon,\lambda,\gamma) \prod_{j=1}^n [1 -
 r_{nj}(1;\epsilon,\lambda,\gamma) - KM] \equiv
\sum_{\nu=0}^n q_{n\nu}(1;\epsilon,\lambda,\gamma) K^\nu M^\nu  \ .
\ee
 This can be compared to the $n$'th order hopping parameter
 expansion of the inverse matrix
\be \label{eq76}
\sum_{\nu=0}^n K^\nu M^\nu = (1-K^{n+1}M^{n+1})/(1-KM) =
(-1)^n \prod_{j=1}^n [e^{2j\pi/(n+1)} - KM] \ .
\ee
 The only change in (\ref{eq75}) with respect to this is the 
 different (optimized) choice of the roots of the $n$'th order
 polynomial.
 From the point of view of the numerical iterative hopping parameter
 expansion \cite{HASMON,REMOPH} this is practically no change in the
 arithmetics, but a considerable gain in precision.
 In addition, there is no convergence problem because for
 $n \to \infty$ we always have $\delta \to 0$.
     
\subsection{An algorithm with optimized hopping parameter expansion}
\label{subsec52}
 The optimized hopping parameter expansion of the previous subsection
 can also be used in the old fermion algorithm based on the direct
 evaluation of the change of the fermion determinant
 \cite{HOPPAR,REMOPH}.
 Let us remember that according to (\ref{eq09}) the change of the
 fermion matrix is
\be  \label{eq77}
\Delta Q_{yv,xu} = - K \left[
\delta_{y,z+\hat{\rho}}\delta_{zx}(1+\gamma_\rho) 
\Delta V_{vu,z\rho} +
\delta_{yz}\delta_{z+\hat{\rho},x}(1-\gamma_\rho) 
\Delta V_{uv,z\rho} \right] \ ,
\ee
 if the link $z\rho$ is changed according to
\be  \label{eq78}
V_{z\rho}^\prime = V_{z\rho} + \Delta V_{z\rho} \ .
\ee
 The ratio of the two fermion determinants is
\be  \label{eq79}
\frac{\det(Q+\Delta Q)}{\det(Q)} = \det (1+\Delta Q \cdot Q^{-1}) \ .
\ee
 The $24 \otimes 24$ matrix needed for the calculation of the
 determinant on the right hand side can be approximately calculated
 by the iterative numerical hopping parameter expansion, or by the
 improved variant discussed above using optimized solvers for the
 inverse.
                                                                                
\section{Discussion}                                   \label{sec6}
 The local bosonic algorithms for gluinos on the lattice investigated
 in section \ref{sec4} seem suitable for numerical experiments aiming
 at an understanding of Yang Mills theories with massive gluinos.
 The study of the limit when the gluino mass goes to zero should be
 possible and could reveal the character of the $N=1$ supersymmetric
 limit by comparing to the expectations based on previous analytical
 work \cite{AKMRV,CURVEN}.
 In particular, the combination of L\"uscher's local bosonic approach
 \cite{LUSCHER} with the noisy correction method \cite{KENKUT} turned
 out to be rather effective in the performed tests.
 Since this combined algorithm has several algorithmic parameters, the
 final optimization for given lattice sizes and bare parameter values
 is a non-trivial task, which has to be carried out in the particular
 applications.

 The required polynomial approximations are defined in a scheme which
 is different from the usual approach leading to Chebyshev
 polynomials.
 The deviation norm is defined as an integral which is quadratic
 in the polynomial coefficients (see section \ref{sec3}).
 This allows a great flexibility in choosing weight factors and
 approximation regions.
 The optimal approximation in this average sense is well suited for
 the present purposes, in particular, in the two-step approximation
 scheme applied in section \ref{sec4}.
 
 Besides the local bosonic algorithms, other applications of the
 optimized polynomial approximation technique of section \ref{sec3}
 have also been briefly discussed.
 In particular, the use of optimized solvers for matrix inversion and
 the application of the optimized hopping parameter expansion
 for a fermion algorithm without pseudofermion fields have been
 considered in section \ref{sec5}.
 The test of this latter algorithm goes beyond the scope of the
 present paper.

 An important question in these algorithms is the r\^ole of the
 deviation norm $\delta$, which is characterizing the quality of the
 polynomial approximations.
 Since exact results are obtained only in the limit $\delta \to 0$,
 this could necessitate an extrapolation of the expectation values
 to $\delta=0$.
 Nevertheless, in the two-step approximation scheme a relative
 precision with $\delta \simeq 10^{-5} - 10^{-6}$ seems possible
 with reasonable effort, therefore the practical need for an
 extrapolation is not clear.
 For instance, such a precision comes already close to the machine
 precision on 32-bit computers, which is generally considered to
 be enough in present numerical simulations.

 Although the methods used here have been formulated and tested in
 case of the SU(2) gauge theory with a Majorana fermion in the
 adjoint representation, they are also applicable in other quantum
 field theories with fermion fields, as QCD or Higgs-Yukawa models
 including also scalar fields.
 
\vspace{1cm}                                                                    
{\large\bf Acknowledgements } \newline                              
\vspace{3pt}                                                                    
                                                                                
\noindent  I thank Martin L\"uscher and Rainer Sommer for discussions
 and helpful suggestions contributing in an essential way to the 
 content of this paper.


\end{document}